\documentclass[notitlepage,showpacs,preprintnumbers]{revtex4}
\usepackage{amsmath,amssymb}
\usepackage{bm,bbm}
\usepackage[T1]{fontenc}
\usepackage{ae,aecompl}
\usepackage{hyperref}
\usepackage{graphicx}
\providecommand{\tphi}{\tilde{\phi}}

\providecommand{\mns}{\mathrm{MNS}}
\providecommand{\tp}{{\mss{\mathsf{T}}}}
\providecommand{\mss}[1]{\mbox{\scriptsize $#1$}}
\providecommand{\ml}[1]{\mbox{\large $#1$}}
\providecommand{\id}{{\mathbbm{1}}}

\DeclareMathOperator{\diag}{\mathrm{diag}}
\DeclareMathOperator{\Tr}{\mathrm{Tr}}
\DeclareMathOperator{\im}{\mathrm{Im}}
\providecommand{\eq}[1]{\begin{equation} #1 \end{equation}}
\providecommand{\eqali}[1]{\begin{equation}\begin{aligned} #1
    \end{aligned}\end{equation}}
\providecommand{\alig}[1]{\begin{align} #1 \end{align}}
\providecommand{\bs}[1]{\boldsymbol{#1}}
\providecommand{\aver}[1]{\langle #1 \rangle}
\providecommand{\ums}[2][1]{\ml{\tfrac{#1}{#2}}}
\providecommand{\xlink}[1]
  {\href{http://arxiv.org/abs/#1}{arXiv:#1}}

\begin{document}
%%%%%%%%%%%%%%%%%%%%%%%%%%%%%%%%%%%%%%%%%%%%%
%\preprint{ArXiv:yymm.nnnn}
\title{An $S_3$ Model for Lepton Mass Matrices with Nearly Minimal Texture}
\author{A.~G.~Dias}%
\email{alex.dias@ufabc.edu.br}
\author{A.~C.~B.~Machado}%
\email{a.c.b.machado1@gmail.com}
\author{C.~C.~Nishi}%
\email{celso.nishi@ufabc.edu.br}
\affiliation{
Universidade Federal do ABC - UFABC,
Santo André, SP, Brasil
}
\date{September 26, 2012}
%%%%%%%%%%%%%%%%%%%%%%%%%%%%%%%%%%
\begin{abstract}
We propose a simple extension of the electroweak standard model based on the
discrete $S_3$ symmetry that is capable of realizing a nearly minimal Fritzsch-type
texture for the Dirac mass matrices of both charged leptons and neutrinos. This is
achieved with the aid of additional $Z_5$ and $Z_3$ symmetries, one of which can be
embedded in $U(1)_{B-L}$. Five complex scalar singlet fields are introduced in
addition to the SM with right-handed neutrinos. Although more general, the modified
texture of the model retains the successful features of the minimal texture without
fine-tuning; namely, it accommodates the masses and mixing of the leptonic sector
and relates the emergence of large leptonic mixing angles with the seesaw mechanism.
For large deviations of the minimal texture, both quasidegenerate spectrum or
inverted hierarchy are allowed for neutrino masses.
\end{abstract}

% 11.30.Hv: Flavor symmetries
% 12.15.Ff: Quark and lepton masses and mixing (see also 14.60.Pq Neutrino mass
% and mixing)
% 12.15.Hh: Determination of Cabibbo-Kobayashi & Maskawa (CKM) matrix elements

\maketitle
%%%%%%%%%%%%%%%%%%%%%%%%%%%%%%%%%%
\section{Introduction}
\label{sec:intro}

As experimental efforts improve our knowledge about the masses and the mixing
pattern of neutrinos, the small neutrino masses and the large mixing angles still
await a natural explanation. The former is successfully accommodated by the
seesaw mechanism but its testability is usually out of reach of our present
experiments. As for the mixing angles, the puzzle is to explain the large angles and
the great difference from the CKM mixing for quarks, which is governed by small
mixing angles.

Concerning the quark sector a scheme relating small mass ratios with small
mixing angles can be devised by assuming a simple texture for the quark mass
matrix\,\cite{Weinberg}.
More precisely, within two families, a hermitean mass matrix with a vanishing (1,1)
element yields the correct mixing angle $\theta_C\sim (m_d/m_s)^{1/2}$.
An extension to three families was proposed by Fritzsch assuming a
minimal texture of the $3\times 3$ matrix, with vanishing (1,1), (1,3), (3,1), (2,2)
elements and a hermitean form\,\cite{Fritzsch}.
Such a texture determines the elements of the CKM matrix as functions of the ratios
of the quark masses.
This minimal texture, however, does not accommodate the present data of quark masses
and mixing structure.

For the lepton sector, it was shown that it is possible to obtain the neutrino
masses and mixing with the same Fritzsch-type texture described above, with the
hermitean form replaced by a symmetric form\,\cite{yanagida}
(see also \cite{xing.02}).
The texture is applied to the Dirac mass matrices for charged leptons and neutrinos
but light neutrino masses arise from the seesaw mechanism.
Consequently, light neutrino masses depend \textit{quadratically} on the
Dirac mass matrix.
This quadratic dependence, in turn, determines that the elements of the PMNS matrix
depend on the ratio of charged lepton masses but on the \textit{square root}
of the ratio of neutrino masses.
This property is what enables large mixing angles to emerge from moderately
hierarchical neutrino masses within this Fritzsch-type texture\,\cite{yanagida}.

More recently, using this minimal texture, Ref.\,\cite{yanagida.12} succeeded
in predicting all observables in neutrino sector from the known values of the
squared-mass differences and mixing angles.  The recently observed
$\theta_{13}$\,\cite{dayabay} was predicted in Ref.\,\cite[b]{yanagida} in the right
range. This information could then be used in Ref.\,\cite{yanagida.12} to make more
precise predictions for the effective mass of double beta decay and the CP violation
measure. There is no ambiguity in hierarchy since the texture only allows the normal
hierarchy for the neutrino masses, excluding either inverse hierarchy or
quasidegenerate masses.

It is interesting to note that the Fritzsch-type texture proposed in
Ref.\,\cite{yanagida} is a particular example of textures with vanishing matrix
elements (\textit{texture zeros}) which was extensively studied in the context of
neutrino mixing; \textit{e.g.}, see Ref.\,\cite{fritzsch.11} for a detailed study
on two-zero textures. Usually, these special textures are supposed to be apparent
for the neutrino mass matrix in the basis where the charged lepton mass matrix is
diagonal\,\cite{frampton}. However, the case in \cite{yanagida} is different, once
the texture proposed is shared by the Dirac mass matrices of both charged leptons
and neutrinos.

On the other hand, from the theoretical point of view, the phenomenologically
successful textures should originate as a consequence of an underlying exact or
approximate flavor symmetry acting at energies above the electroweak scale. For
example, abelian symmetries can be systematically used to justify texture
zeros\,\cite{grimus:texture}.
More specifically, one can obtain the necessary texture zeros of the Fritzsch
ansatz for quarks from a $Z_4$ symmetry within a 2-Higgs-doublet
model\,\cite{branco:z4}.
Some further relations between the nonzero entries of the mass matrices may arise
from an underlying unifying gauge theory, which naturally accommodates a symmetric
or
hermitean mass matrix such as in $SU(5)$ constructions\,\cite{simoes.12} (symmetric
$M_u$) or the original left-right proposal\,\cite{Fritzsch} (hermitean $M_{u,d}$).

In contrast to abelian symmetries, the use of discrete nonabelian symmetries is
particularly appealing\,\cite{ishimori} to generate \textit{mass-independent} mixing
patterns\,\cite{lam:texture}, the most interesting of which is the tribimaximal
mixing. In this context, models using the simple $S_3$ symmetry can be found
abundantly in the literature\,\cite{s3}.

In this work, we take a different perspective and use the nonabelian $S_3$ flavor
symmetry, together with some abelian symmetries, to impose texture zeros
and, at the same time, relate some nonzero elements of the mass matrices.
The latter can not be accomplished within the gauge structure of the SM if we only
impose abelian symmetries.
The resulting form for the mass matrices is a modified form of the ansatz of
Ref\,\cite{yanagida}.
The texture for the mass matrices of charged leptons and neutrinos
has in fact a common origin according to our proposal.

Other approaches to the mass and mixing problem as we deal with here can be found in
the literature as, \textit{e.g}, the use of  $S_3$ symmetry aiming at
explaining a Fritzsch-type texture\,\cite{meloni:s3}.
The latter, however, is obtained only for the neutrino mass matrix \textit{itself};
\textit{videlicet} a seesaw mechanism is not considered and the charged lepton mass
matrix is also not of the Fritzsch-type but nearly diagonal.

The outline of the paper is the following. In Sec.\,\ref{sec:minimal}, we review the
minimal texture hypothesis of Ref.\,\cite{yanagida}. In Sec.\,\ref{sec:model}, we
present the model. We show in Sec.\,\ref{sec:modified} an analysis of the
modified form for the ansatz and some numerical examples.
The conclusions are shown in Sec.\,\ref{sec:conclusions}.
In appendix \ref{ap:pot}, we discuss the scalar potential and justify the
necessary alignment for the vevs of some scalar fields.

%%%%%%%%%%%%%%%%%%%%%%%%%%%%%%%%%%%%%%
\section{The minimal texture ansatz}
\label{sec:minimal}

The minimal texture hypothesis of Refs.\,\cite{yanagida} and
\cite{yanagida.12} consists of
assuming that the mass matrix for charged leptons and the Dirac mass matrix for
neutrinos have the simple
form\,\cite{Fritzsch}
\eq{
\label{fritzsch}
M=
\left(\begin{array}{ccc}
0 & A & 0 \\
A & 0 & B \\
0 & B & C
\end{array}\right)\,,
}
where $A,B,C$ are complex numbers.

With a rephasing of lepton fields we can transform \eqref{fritzsch} to a real
symmetric matrix
\eq{
\label{fritzsch:real}
|M|=
\begin{pmatrix}
0 & |A| & 0 \\
|A| & 0 & |B| \\
0 & |B| & |C|
\end{pmatrix}\,.
}
We will denote the matrix $|M|$ in \eqref{fritzsch:real} as the \textit{real form}
of the matrix $M$ in \eqref{fritzsch}. The matrix in the real form can be
diagonalized by a real orthogonal
matrix. It follows that $M$ can be diagonalized by
\eq{
\label{ulmur}
M^{diag}=U^\tp M U\,,
}
where
\eq{
\label{Uger:l}
U=d^\dag O\,,
}
$d$ is a diagonal rephasing matrix and $O$ a real orthogonal matrix (the
transformation in (\ref{ulmur}) is a special case of a biunitary transformation).
The three parameters $|A|,|B|,|C|$ are completely determined by the three
eigenvalues of $M$, $(m_1,-m_2,m_3)$, where $0\le m_1< m_2< m_3$.
When $m_1\ll m_2 \ll m_3$, this ansatz implements the idea that small mixing
angles are consequences of large hierarchies in mass.

Let us attribute the simple form \eqref{fritzsch} for the mass matrix of charged
leptons and for the Dirac mass matrix for neutrinos,
\eq{
\label{Mlnu:yana}
M_l =  \left(\begin{array}{ccc}
0 & A_l & 0  \\
A_l & 0   &  B_l \\
0 &  B_l &  C_l
\end{array}\right),\quad
M_D = \left(\begin{array}{ccc}
0 & A_{\nu} & 0  \\
A_{\nu} & 0   &  B_{\nu} \\
0 &  B_{\nu} &  C_{\nu}
\end{array}\right).
}
with both being diagonalized by matrices of the form \eqref{Uger:l}.

It is assumed here that  the light neutrino masses are generated through the
seesaw mechanism from the mass matrix
\eq{
\label{seesaw}
M_\nu= -M_D^\tp M_R^{-1}M_D\,,
}
where $M_R$ is the right-handed Majorana neutrinos mass matrix. For simplicity, it
is further assumed that
$M_R$, in the basis where $M_D$ is diagonal, is proportional to the identity
\,\cite[b]{yanagida}, i. e.,
\begin{eqnarray}
U_{\nu }^\tp M_R U_{\nu } = M_0\id_3\,
\end{eqnarray}
$M_0$ is real, positive and much larger than the Dirac neutrino mass scale.
This means that, in the basis where $M_D$ has the form \eqref{Mlnu:yana},
\eq{
\label{MR:yana}
M_R=M_0d_{\nu }^{ 2}\,,
}
where $d_\nu$ is the rephasing matrix for $M_D$.
In this case, the mass matrix $M_\nu$ is still diagonalized by
\eq{
U_{\nu }=d_{\nu }^\dag O_\nu\,,
}
\eq{
\label{Mnudi}
M_\nu^{diag}=-\frac{1}{M_0}[M_D^{diag}]^2=U_{\nu }^\tp M_\nu U_{\nu }\,,
}
i.e., the same matrix that diagonalizes $M_D$.
This conclusion is not significantly modified for a $M_R$ different from
\eqref{MR:yana} if the eigenvalues of $M_D$ are hierarchical\,\cite[b]{yanagida};
in particular, the effects of additional phases in $M_R$ are only
minor\,\cite[a]{yanagida}.

The PMNS matrix is then given by
\eq{
\label{MNS}
V_{\mns}=U_l^\dag U_\nu=O_l^\tp Q O_\nu\,,
}
where $Q=\diag(1,e^{i\sigma},e^{i\tau})$ is a matrix of phases coming from $d_l
d_\nu^\dag$ subtracting a global phase.

With this minimal texture, Ref.\,\cite{yanagida.12} succeeds in predicting all the
relevant observables in the neutrino sector, such as the Jarlskog invariant or the
absolute mass scale, from the presently known mass-squared differences and mixing
angles.
Our main interest in this paper is to propose a model realization of this
ansatz.

%%%%%%%%%%%%%%%%%%%%%%%%%%%%%%%%
\section{The model}
\label{sec:model}

We propose a model based on $S_3$ symmetry that is capable of naturally
generating the simple form \eqref{fritzsch} for the Dirac mass matrices for charged
leptons and neutrinos.

We enlarge the symmetry group of the SM by including a flavor group $G_F=S_3\otimes
Z_5\otimes Z_3$.
The three families of left-handed lepton doublets $L_i$ and right-handed singlets
$l_{iR}$ and $\nu_{iR}$, $i=1,2,3$ (or $l_i=e,\mu,\tau$), transform nontrivially
under $G_F$.
The Higgs doublet $\phi$ is a singlet of $G_F$.
We also assume that the $G_F$ symmetry is valid only on a scale above the
electroweak scale, where new physics effects can be described by some
nonrenormalizable $G_F$-symmetric interactions which depend on five complex scalars
 $\eta$,  $\chi$, $\chi'$, and $\zeta_i$, $i=1,2$, that are complete singlets of the
SM.

We arrange the fields of the model on multiplets transforming under irreducible
representations of $G_F$. There are only three such representations of $S_3$: two
singlets and a doublet, which are denoted as $\bs{1},\bs{1'},\bs{2}$. We assign
them as
\eqali{
\bs{2}:&~L_{D} \equiv
( L_e, L_\mu),\,
    E_{D} \equiv
    ( e_R, \mu_R),\,
\cr&~
    N_{D} \equiv
    (\nu_{1R}, \nu_{2 R}),\,
    \zeta_D\equiv
    (\zeta_{1}, \zeta_{2});\,
\cr
\bs{1}:&~L_S\equiv L_\tau \,,\; E_{S} \equiv \tau_R \,,\;
    N_{S} \equiv \nu_{3 R},
  \cr&~\chi\,,\;\chi'\,;
\cr
\bs{1}':&~\eta\,.
}
The complete assignment of representations of $G_F$ is shown in Table
\ref{tab:irreps}.

\begin{table}[ht]\centering
\vspace{-\baselineskip}
\eq{\nonumber
\begin{array}{|c|c|c|c|c|c|c|c|c|c|c|c|}
\hline\rule[0cm]{0cm}{.9em}
  & L_D & L_S & E_D & E_S & N_D & N_S & \phi  &
\zeta_D & \eta & \chi & \chi' \\[-.1ex]
\hline\rule[0cm]{0cm}{.9em}
 S_3 & 2 & 1 & 2 & 1 & 2 & 1 & 1 & 2 & 1'& 1 & 1\\
\hline\rule[0cm]{0cm}{1em}
 Z_5 & \omega_5^4 & 1 & \omega_5 & 1 & \omega_5 & 1 & 1  & \omega_5^4 & \omega_5^2 &
1 &\omega_5^3\\[-.1ex]
\hline\rule[0cm]{0cm}{1em}
 Z_3 & \omega  & \omega & \omega  & \omega & \omega & \omega  & 1  & 1
 &1 & \omega & \omega\\[-.1ex] \hline
\end{array}
}\vspace*{-1\baselineskip}
  \caption{Transformation properties under $G_F=S_3{\otimes} Z_5{\otimes} Z_3$
where $\omega=e^{i2\pi/3}$ and $\omega_5=e^{i2\pi/5}$.}
\label{tab:irreps}
\end{table}

The relevant branching rule for $S_3$ is
$
\textbf{2}\otimes \textbf{2}=\textbf{1}\oplus\textbf{1}^\prime\oplus \textbf{2}\,.
$
For two doublets $x=(x_1,x_2)^\tp$ and $y=(y_1,y_2)^\tp$, the decomposition
can be performed explicitly as\,\cite{ishimori}
\eqali{~
[x\times y]_1&=x_1y_1+x_2y_2\,,\quad
[x\times y]_{1'}=x_1y_2-x_2y_1\,,\quad\cr
[x\times y]_{2}&=
\begin{pmatrix} x_2 y_2 - x_1 y_1 \\ x_1 y_2 + x_2 y_1 \end{pmatrix} .
\label{ishimorinote}
}
Notice the following products for three doublets $x,y,z$:
\eqali{
\label{222=1}
[x\times y\times z]_1&\equiv
[[x\times y]_2\times z]_1=[x\times [y\times z]_2]_1
\,,\cr
[x\times y\times z]_{1'}&\equiv
[[x\times y]_2\times z]_{1'}= -[x\times [y\times z]_2]_{1'}
\,,
}
are uniquely defined.

We can now easily write the Yukawa Lagrangian for charged leptons,
\begin{eqnarray}
\label{yukawa:l}
-\mathcal{L}^{Y}_{l}&=&
  \frac{a_l^*}{\Lambda}[\bar{L}_{D} \phi E_{D}]_{1'}\,\eta^*
  + \frac{{b'_l}^*}{\Lambda}\bar{L}_S\phi[E_{D}\zeta_D]_1
  \cr&&
  +\ \frac{b_l^*}{\Lambda}[\bar{L}_D\zeta_D]_1 \phi E_S
  + c_l^*\bar{L}_S\phi E_S
  +h.c.,
\end{eqnarray}
where we have used \eqref{222=1} and only retained operators of order up to
$1/\Lambda$\,\cite{endnote0}.
Notice the tau lepton is the only one that receives mass through renormalizable
interactions, hence its large mass.
In the basis $\bar{l}_{iR}l_{jL}$, we obtain the mass matrix
\begin{eqnarray}
\label{Ml}
M_l &=&  \left(\begin{array}{ccc}
0 & -A_l & 0  \\
A_l & 0   &  B_l' \\
0 &  B_l &  C_l
\end{array}\right)\,,
\end{eqnarray}
where the elements are
\begin{eqnarray}
A_l = a_lv_\phi \frac{u_\eta}{\Lambda}\,,~
B_l = b_l v_\phi \frac{u_2^*}{\Lambda}\,,~
B'_l = {b'_l} v_\phi \frac{u_2^*}{\Lambda}\,,~
C_l = c_lv_\phi\,.
\end{eqnarray}
These elements depend on the vacuum expectation values (vevs) of the neutral fields
which we assume have the form
\eq{
\label{vevs}
\aver{\zeta_D}=(0,u_2)^\tp\,,~
\aver{\eta}=u_\eta\,,~
\aver{\phi^0}=v_\phi\,,~
}
where $\sqrt{2}v_\phi=v=246\rm GeV$ is the electroweak scale; $u_2$ and $u_\eta$ may
be complex a priori.
Notice the minus sign in the (12) entry of \eqref{Ml} can be eliminated by
rephasing the appropriate right-handed lepton field.

Analogously, the Dirac mass matrix for neutrinos is generated by the effective
Yukawa
Lagrangian
\begin{eqnarray}
- \mathcal{L}^{Y}_{\nu}&=&
  \frac{a_\nu^*}{\Lambda} [\bar{L}_{D}\tphi N_{D}]_{1'}\,\eta^*
  +\frac{{b'_{\nu}}^*}{\Lambda}\bar{L}_S \tphi [N_{D}\zeta_D]_1
\cr&&
  +\ \frac{b_\nu^*}{\Lambda}[\bar{L}_D\zeta_D]_1 \tphi N_{S}
  + c_\nu^*\bar{L}_S \tphi N_{S}
  +h.c.
\label{yukawa:nu}
\end{eqnarray}
By assuming the same vevs \eqref{vevs} we obtain the same form as
\eqref{Ml},
\begin{eqnarray}
\label{MD}
M_D &=&  \left(\begin{array}{ccc}
0 & -A_\nu & 0  \\
A_\nu & 0   &  B_\nu' \\
0 &  B_\nu &  C_\nu
\end{array}\right)\,,
\end{eqnarray}
with the identification
\begin{equation}
A_\nu = a_\nu v_\phi \frac{u_\eta}{\Lambda}\,,~
B_\nu = b_\nu v_\phi \frac{u_2^*}{\Lambda}\,,~
B'_\nu = {b'_\nu} v_\phi\frac{u_2^*}{\Lambda}\,,~
C_\nu = c_\nu v_\phi\,.
\end{equation}

The Majorana mass terms are generated by
\begin{eqnarray}
\label{L:MR}
- \mathcal{L}^{M}_{\nu}&=&
  \ums{2}\lambda_1[\bar{N}_{D} N^c_{D}]_1\chi^{\prime*}
  +\ums{2}\lambda_2\bar{N}_{S} N^c_{S}\chi^*
  +h.c.
\end{eqnarray}
These terms generate the Majorana mass matrix
\begin{eqnarray}
\label{MR}
M_R &=&  \diag(\mu_1,\mu_1,\mu_2)\,,
\end{eqnarray}
where $\mu_1=\lambda_1{u_S'}^*,\,\mu_2=\lambda_2{u_S}^*$ and
\eq{
\label{vevsz}
\aver{\chi'}=u'_S\,,~
\aver{\chi}=u_S\,.
}
The vevs $u_S,u_S'$ may be complex.

The mass matrix for light neutrinos is given by the seesaw formula \eqref{seesaw},
which leads to
\eq{
\label{Mnu}
M_\nu= -
\begin{pmatrix}
\frac{A^2_\nu}{\mu_1} & 0 & \frac{A_\nu B'_\nu}{\mu_1}\cr
0 & \frac{A^2_\nu}{\mu_1}+\frac{B^2_\nu}{\mu_2} & \frac{B_\nu C_\nu}{\mu_2}\cr
\frac{A_\nu B'_\nu}{\mu_1} & \frac{B_\nu C_\nu}{\mu_2} &
\frac{{B'_\nu}^2}{\mu_1}+\frac{C^2_\nu}{\mu_2}
\end{pmatrix}\,.
}

From Eqs.\,\eqref{Ml}, \eqref{MD} and \eqref{MR}, we recover the minimal texture of
Ref.\,\cite{yanagida} if
(i) $|\mu_1|=|\mu_2|=M_0$ in $M_R$
[and $M_R=M_0\id_3$ in the basis where $M_D$ has the real form
\eqref{fritzsch:real}],
(ii) $B_l'=B_l$ in $M_l$, and
(iii) $B_\nu=B_\nu'$ in $M_D$.
It is argued in Ref.\,\cite{yanagida} that (i) is not essential as
long as $M_D$ has hierarchical eigenvalues.
We can confirm this by applying a rephasing transformation to \eqref{Mnu} to
obtain
\eq{
\label{Mnu:real}
d_\nu^\dag M_\nu d_\nu^\dag=
\begin{pmatrix}
a^2e^{-i2\delta_2} & 0 & ab'\cr
0 & a^2e^{i2\delta_1}+b^2 & bc\cr
ab' & bc & {b'}^2e^{i2\delta_2}+c^2
\end{pmatrix}\,.
}
We have used the shorthands for the positive real numbers
$a^2=|A^2_\nu/\mu_1|$, $b^2=|B^2_\nu/\mu_2|$, ${b'}^2=|{B'_\nu}^2/\mu_1|$,
$c^2=|C^2_\nu/\mu_2|$
and
for the phases $\delta_1=\arg(A_\nu B^*_\nu/\sqrt{\mu_1\mu_2^*})$,
$\delta_2=\arg({B'_\nu}C^*_\nu/\sqrt{\mu_1\mu_2^*})$.
Given the hierarchy $c\gg b,b' \gg a$, we can see that each element of $M_\nu^\dag
M_\nu$, calculated from \eqref{Mnu:real}, does not depend on the phases
$\delta_1,\delta_2$ in the leading terms.
Thus we can consider $M_\nu$ to be real in the first approximation.

We will show in the next section how (ii) and (iii) affect the relations between
the parameters of $M_l,M_\nu$ and the masses.

%%%%%%%%%%%%%%%%%%%%%%%%%%%%%%%%%%%%%%%%%%%%%%%%%%%%%%%%%%%%%%%%
\section{Deviations from the minimal texture}
\label{sec:modified}

Let us analyze the consequences of the deviation of $M_l$ and $M_D$ from the
minimal texture \eqref{fritzsch}.

The texture obtained for $M_l$ and $M_D$ in our model, Eqs.\,\eqref{Ml}
and \eqref{MD}, has the modified form
\eq{
\label{M:mod}
M'=\begin{pmatrix}
0 & -A & 0\cr
A & 0 & B'\cr
0 & B & C
\end{pmatrix}\,.
}
It generalizes the minimal texture \eqref{fritzsch} in that $B'$ and $B$ are not
necessarily equal. We expect, however, that $|B|\sim |B'|$ such that the hierarchy
$|A|\ll |B|,|B'|\ll |C|$ is maintained in conformity to the hierarchy of
masses $m_1\ll m_2\ll m_3$.
If we also had the elements (1,2) and (2,1) distinct in \eqref{M:mod}, we
would obtain the nearest-neighbor-interaction (NNI) form\,\cite{branco.mota}, which
is always achievable by a weak basis change.

Since $M^\prime$ in \eqref{M:mod} is no longer symmetric, even if we eliminate the
minus sign in the (12) entry, the relevant mass matrix that furnishes the (squared)
masses is $M^{\prime \dag} M'$.
This matrix can still be transformed to the real form
\eq{
\label{MM:mod:r}
|M^{\prime\dag}M'|=
\begin{pmatrix}
a^2 & 0 & b'a\cr
0 & a^2+b^2 & bc\cr
b'a & bc & {b'}^2+c^2
\end{pmatrix}\,
}
by rephasing the appropriate fields.
We have also used the shorthands $a\equiv |A|$, $b\equiv|B|$, $b'\equiv |B'|$,
$c\equiv|C|$.
Note that \eqref{Mnu:real} has the same form as \eqref{MM:mod:r} when the phases
are neglected.

The characteristic equation for \eqref{MM:mod:r} is
\eq{
\label{MM:mod:eq}
\lambda^3-(2a^2+b^2+{b'}^2+c^2)\lambda^2+\big[(a^2+b^2)(a^2+{b'}^2)+2a^2c^2\big]
\lambda-a^4c^2=0\,.
}
This equation should be compared to the characteristic equation for $|M|^2$ which is
obtained from \eqref{MM:mod:eq} when $b'=b$.
We know $|M|^2$ has eigenvalues $(m_1^2,m_2^2,m_3^2)$ and the same eigenvectors of
$|M|$ in \eqref{fritzsch:real}.
If we also identify the eigenvalues of $|M^{\prime\dag}M'|$ as
$(m_1^2,m_2^2,m_3^2)$,
we can still write $a,c$ and $\bar{b}\equiv\sqrt{\ums{2}(b^2+{b'}^2)}$ as
functions of the masses and one remaining degree of freedom, quantified by
$\Delta b^2\equiv b^2-{b'}^2$.
We assume $m_3>m_2>m_1\ge 0$ for the expressions below.

Let us analyze the relations of $a,c,\bar{b}$ with the masses and $\Delta b^2$.
The relation between $a$ and $c$ is the same as in the minimal texture, i.e.,
\eq{
\label{eq:a}
a=\sqrt{\frac{m_1m_2m_3}{c}}\,,
}
which follows from $\det(|M^{\prime\dag}M'|)=m_1^2 m_2^2 m_3^2$.
The relation between $\bar{b}$ and $c$ is given by
\eq{
\label{eq:bbar}
(\bar{b})^2=\ums{2}(m_1^2+m_2^2+m_3^2-c^2)-\frac{m_1m_2m_3}{c}\,,
}
which follows from  $\Tr(|M^{\prime\dag}M'|)=m_1^2 +m_2^2+ m_3^2$.
At last, the parameter $c$ can be obtained as a root of
\eq{
\label{eq:c}
(c^2-m_1^2-m_2^2-m_3^2)^2-(\Delta b^2)^2=
4(m_1^2m_2^2+m_2^2m_3^2+m_1^2m_3^2)-8m_1m_2m_3c\,.
}
Once $c$ is fixed by \eqref{eq:c}, $a$ and $\bar{b}$ are known for a given $\Delta
b^2$.

To obtain $c$, we should analyze Eq.\,\eqref{eq:c}. Among the possible multiple
roots, we identify the physical root as the one that reduces to the known
expression\,\cite{Fritzsch}
\eq{
c_0\equiv m_1-m_2+m_3\,,
}
in the limit $\Delta b^2\to 0$.
In general, we should write
\eq{
c=c_0+\delta c\,.
}
To quantify the deviation $\delta c$, we should rewrite Eq.\,\eqref{eq:c} in terms
of the relative deviation $\delta c/c_0$, which gives
\eq{
\label{eq:dc}
\ums{4}\bigg(\frac{\delta c}{c_0}\bigg)^4+\bigg(\frac{\delta c}{c_0}\bigg)^3
+\alpha_2\bigg(\frac{\delta c}{c_0}\bigg)^2+\alpha_1\bigg(\frac{\delta
c}{c_0}\bigg)=
\bigg(\frac{\Delta b^2}{2c_0^2}\bigg)^2\,.
}
We have used the shorthands
\eqali{
\label{alpha:12}
\alpha_1&= -2\frac{b_0^2}{c_0^2}\,,\cr
\alpha_2&= \frac{m_1^2+m_2^2+m_3^2+3[m_1m_3-m_2(m_1+m_3)]}{c_0^2}
\,,
}
and
\eq{
\label{def:b0}
b_0\equiv \sqrt{\frac{(m_2-m_1)(m_1+m_3)(m_3-m_2)}{m_1-m_2+m_3}}\,.
}
The latter corresponds to $b=b'$ in the minimal texture and can be obtained
from \eqref{eq:bbar} in the limit $c=c_0$.
We can confirm from expression \eqref{def:b0} that indeed the negative eigenvalue of
\eqref{fritzsch:real} should be associated to the intermediate mass.

Equation \eqref{eq:dc} can be solved numerically but we can seek an approximate
solution that goes to zero as $\Delta b^2\to 0$.
To quantify $\Delta b^2$ better, we can parametrize $\Delta b^2$ by either
\eqali{
\label{def:beta}
\Delta b^2&=2\bar{b}^{\,2}\beta\,~~\text{ or }\cr
\Delta b^2&=2b_0^2\beta_0\,,
}
depending on the choice of using the parameters $\{a,\bar{b},c\}$ or the
masses $\{m_i\}$ as input parameters.
Since $b,b'$ approach $b_0$ in the limit $\Delta b^2\to 0$,
$\beta\approx\beta_0$ for small values.
For large values, $\beta$ strictly obeys $|\beta|\le 1$ by definition whereas
$|\beta_0|$ can assume values larger than unity but we still need $|\beta|\lesssim
1$ to ensure $b\sim b'$\,\cite{endnote1}.
By noting from \eqref{alpha:12} that $|\alpha_1|\ll 1$ whereas
$\alpha_2\sim\mathcal{O}(1)$, we can drop the cubic and quartic term in
\eqref{eq:dc} and write an approximate solution as
\eq{
\label{dc:approx}
\frac{\delta c}{c_0}\approx
-\frac{b_0^2}{c_0^2}
f(\beta_0)\,,
}
where
\eq{
\label{def:f}
f(\beta_0)\equiv
\frac{1}{\alpha_2}\Big[\sqrt{1+\alpha_2\beta_0^2}-1\Big]
\,.
}
For a hierarchical spectrum $b_0^2/c_0^2\approx m_2/m_3$, and the error for
\eqref{dc:approx} is of the order of $\beta_0^6b_0^6/c_0^6\sim
\beta_0^6m_2^3/m_3^3$.
We can see that the relative deviation $\delta c/c_0$ is small and $\delta c$ is
at most of the order of $m_2$ as long as $b\sim b'$.
We can also see from \eqref{eq:a} that the relative deviation $\delta a/a_0$ is
small and of the order of \eqref{dc:approx}.

As $b,b'$ are expected to be of the same order, they vary significantly with
$\beta_0$ (or $\beta$) in \eqref{def:beta}. We can write $b,b'$ as functions of the
masses and $\beta_0$.
Thus, for a given $\beta_0$ and masses $\{m_i\}$, the matrix \eqref{MM:mod:r} is
fixed.

We should also comment on the dependence of the eigenvectors of \eqref{MM:mod:r}
on $\beta_0$ as it deviates from 0. Let us denote by $O$ the matrix of
eigenvectors of \eqref{MM:mod:r}, with eigenvectors associated to
$m_1^2,m_2^2,m_3^2$ arranged in the columns 1,2,3, respectively.
For the minimal texture ($\beta_0=0$), $O$ is fixed\,\cite{yanagida}.
With hierarchical masses, the diagonal elements $O_{11},O_{22},O_{33}$ are the
largest in modulus and we use the convention that they are positive;
$O_{31},O_{12},O_{32}$ are thus negative.
As $\beta_0$ varies for $|\beta_0|\lesssim 1$, one can check that the elements
$O_{21},O_{12},O_{13}$ are approximately constant with $\beta_0$ ($|O_{12}|$
increases
mildly and $|O_{13}|$ decreases mildly with $\beta_0$).
On the other hand, $|O_{31}|,|O_{32}|,|O_{23}|$ increase significantly with
$\beta_0$.
The diagonal elements decrease accordingly.
We can see that the mixing angles can be significantly modified.

If we extend this discussion to $M_l$ and $M_D$ of Eqs.\,\eqref{Ml} and \eqref{MD},
their modified texture \eqref{M:mod} introduces two more degrees of freedom that we
can parametrize as $\beta_{0l}$ and $\beta_{0\nu}$.
Given that the charged lepton masses have large hierarchy, one can check that the
elements $(12),(13),(21),(31)$ of $O_l$ are small and do not change significantly
for $|\beta_0|\lesssim 1$; $|(O_l)_{23}|,|(O_l)_{32}|$ increase substantially with
$\beta_{0l}$.
We can conclude that the space of solutions found in Ref.\,\cite{yanagida.12} is
significantly broadened as $\beta_{0l},\beta_{0\nu}$ can be varied in the range
$|\beta_{0\alpha}|\lesssim 1$, $\alpha=l,\nu$.
Besides the solutions close to $\beta_{0l}=\beta_{0\nu}=0$, which are necessarily
present, one can easily find solutions away from that point.

\begin{figure}[h]
\centering
\includegraphics[scale=0.3,angle=90]{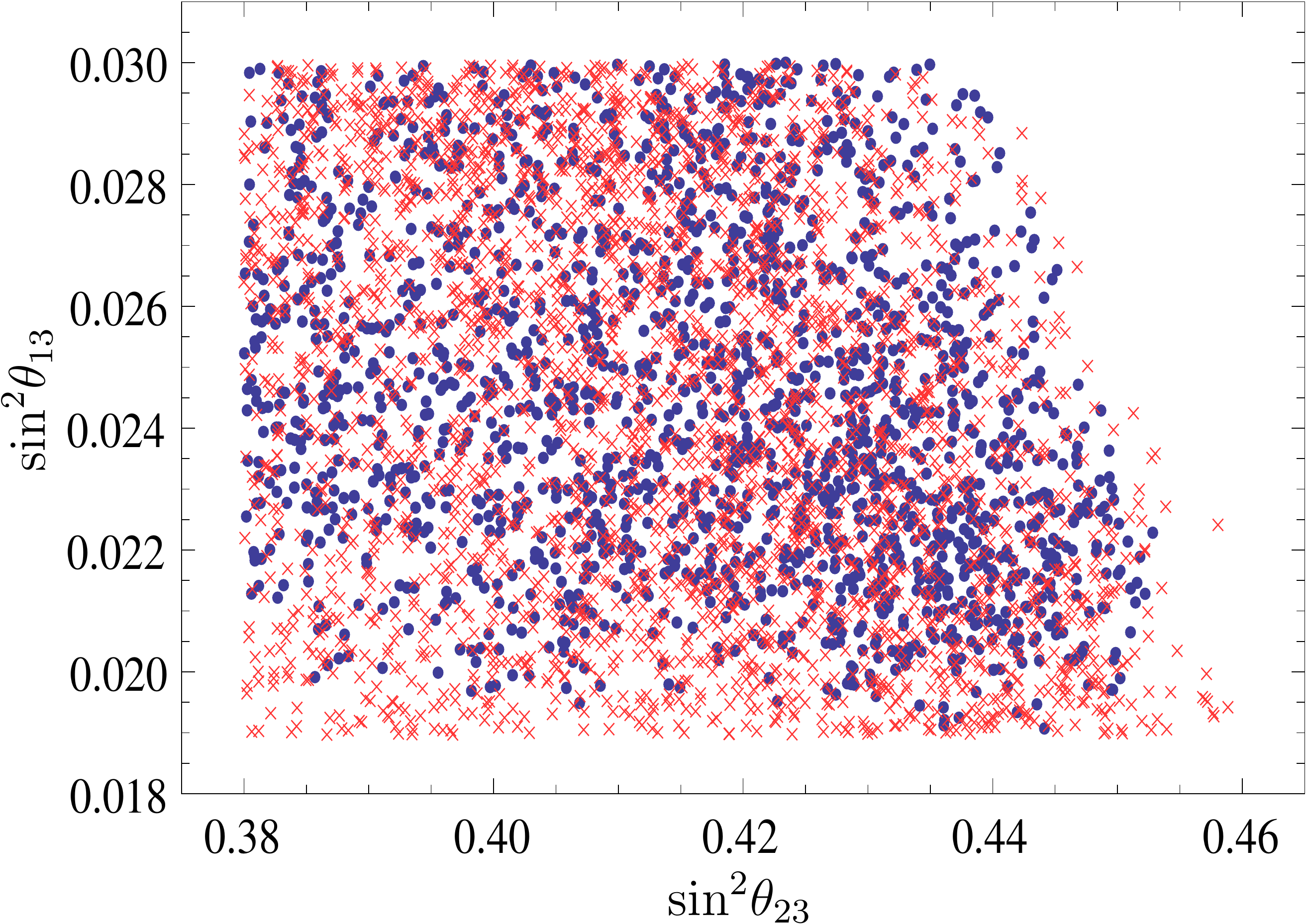}
\caption{$\sin^2\!\theta_{13}$ as a function of $\sin^2\!\theta_{23}$;
the symbols $\times$ correspond to $\beta_{0l}=\beta_\nu=0$ while
dots correspond to $(\beta_{0l},\beta_\nu)=(1,-0.418)$.
}
\label{fig1}
\end{figure}

\begin{figure}[h]
\centering
\includegraphics[scale=0.3,angle=90]{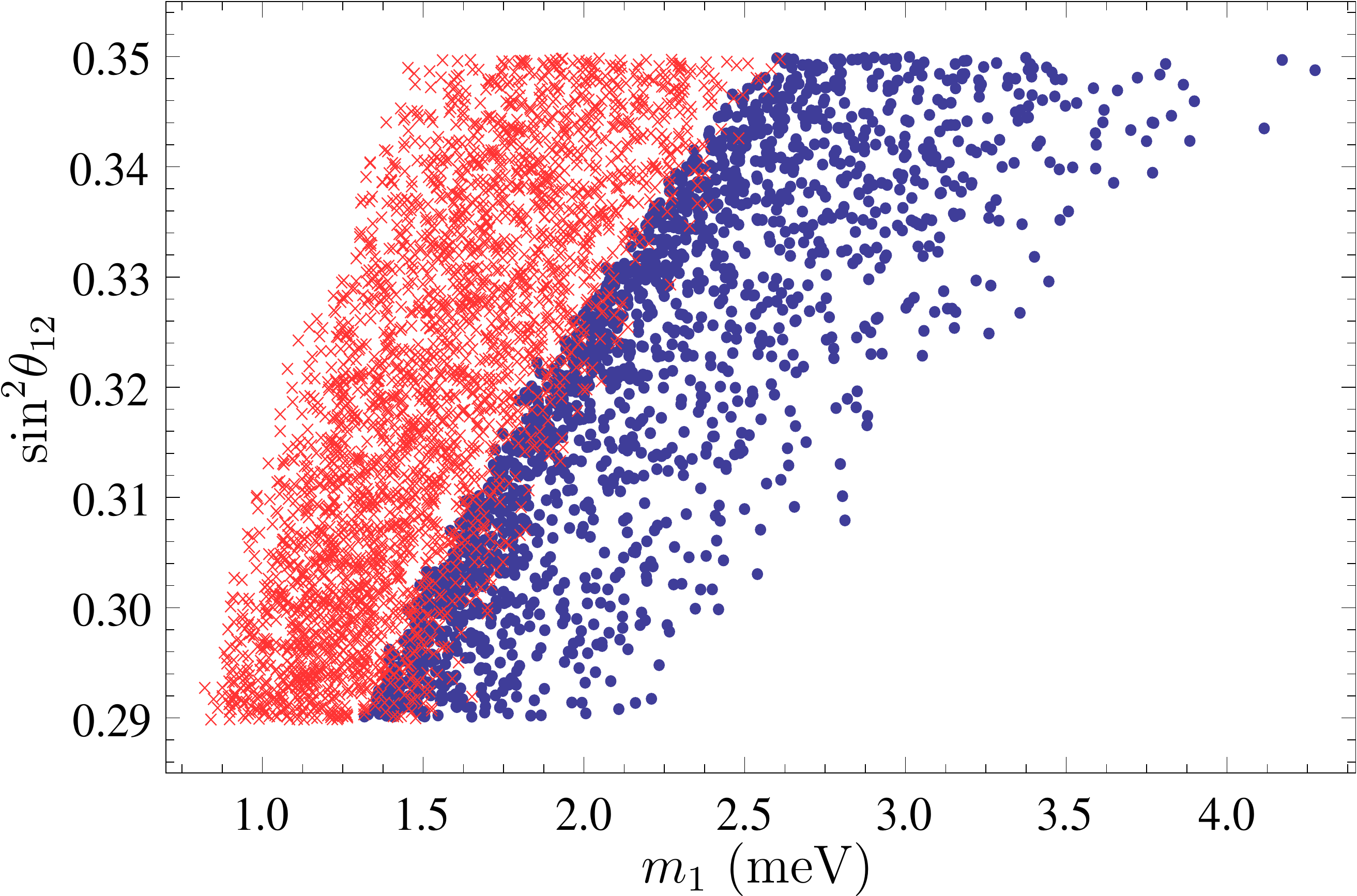}
\caption{$\sin^2\!\theta_{12}$ as a function of the lightest neutrino mass;
the symbols $\times$ correspond to $\beta_{0l}=\beta_\nu=0$ while
dots correspond to $(\beta_{0l},\beta_\nu)=(1,-0.418)$.
}
\label{fig2}
\end{figure}

As an example, we make a numerical comparison between two cases: (a) the minimal
texture ($\beta_{0l}=\beta_\nu=0$) and (b) $(\beta_{0l},\beta_\nu)=(1,-0.418)$;
the latter is substantially different from the minimal case.
It is appropriate to use $\beta_\nu$ instead of $\beta_{0\nu}$ because we use the
parameters $a,\bar{b},c$ as input parameters instead of the neutrino masses.
The comparison is shown in two figures.
In Fig.\,\ref{fig1}, we show $\sin^2\!\theta_{13}$ as a function of
$\sin^2\theta_{23}$. We can see the two cases are indistinguishable.
Fig.\,\ref{fig2} shows $\sin^2\!\theta_{12}$ as a function of the lightest
neutrino mass $m_1$ where the two cases (a) and (b) can be almost completely
separated. We also note that $m_1$ is in general nearly twice as heavy for case (b)
compared to case (a).

To generate our points for the scatter plots, we employ the following
procedure.
We use as inputs for the charged lepton sector, the central values of the masses
$(m_e,m_\mu,m_\tau)$\,\cite{pdg} and $\beta_{0l}$, which determine the mass matrix
\textit{squared} \eqref{MM:mod:r}.
For the neutrino sector, we use $\{a,\bar{b},c,\beta_\nu\}$ as inputs to fix the
neutrino mass matrix \eqref{Mnu:real}, with $\delta_1=\delta_2=0$.
The two phases $\sigma,\tau$ in \eqref{MNS} complete the list of free parameters.
Then, for fixed values of $(\beta_{0l},\beta_\nu)$, we vary the parameters
$\{a,\bar{b},c,\sigma,\tau\}$ randomly and select only the points
compatible, within 2-$\sigma$, with the values of the mixing angles,
$\theta_{12},\theta_{13},\theta_{23},$ and mass-squared differences, $\Delta
m^2_{21}, \Delta m^2_{31}$, of Ref.\,\cite{valle.12}.
No analytical approximations are employed in the diagonalization process.
Although we use slightly different data, we can see our points are compatible with
the predictions of \cite{yanagida.12} for the minimal texture.
For the nonminimal case, we assume normal hierarchy and only points with parameters
$\{a,\bar{b},c\}$ close to the minimal case are sought.

For completeness, we list some typical values for the parameters for the two cases:
\eqali{
\label{typical}
\text{(a)}&\quad
a=0.07\,{\rm eV},~\phantom{83}
\bar{b}=0.105\,{\rm eV},~\phantom{4}
c=0.170\,{\rm eV},~
\sigma=1.5\,,~
\tau=-1.78\,;\cr
\text{(b)}&\quad
a=0.0783\,{\rm eV},~
\bar{b}=0.0984\,{\rm eV},~
c=0.175\,{\rm eV},~
\sigma=2.4\,,~
\tau=-0.65\,.
}
They are compatible within 1-$\sigma$ of Ref.\,\cite{valle.12}.
While for case (a), the phases in \eqref{typical} are really typical, together with
their opposite signs ($\sigma\to -\sigma$ and $\tau\to -\tau$), the phase $\tau$ for
case (b) is well distributed in the whole range $(-\pi,\pi]$.

For an even larger departure from the minimal texture, one can find solutions which
were not possible for the minimal texture\,\cite{yanagida}, namely, neutrino masses
with quasidegenerate spectrum (QD) or inverted hierarchy (IH).
One example of solutions have parameters
\eqali{
\label{qd.ih}
\text{(QD)}&\quad
\beta_{0l}=2.5\,,~
% \beta_l\approx 0.9285
% \beta_{0\nu}\approx -0.42
\beta_{\nu}=-0.41\,,~
a=0.225\,{\rm eV},~
\bar{b}=0.0115\,{\rm eV},~
c=0.265\,{\rm eV},~
\sigma=1.6\,,~
\tau=-1.8\,;\cr
\text{(IH)}&\quad
\beta_{0l}=5.2\,,~
% \beta_l=0.982027
% \beta_{0\nu}=0.10031
\beta_{\nu}=0.107\,,~
a=0.220\,{\rm eV},~
\bar{b}=0.0349\,{\rm eV},~
c=0.0608\,{\rm eV},~
\sigma=1.6\,,~
\tau=-1.8\,,
}
which lead to the neutrino masses 
\eqali{
(m_1,m_2,m_3)_{\rm QD}&= (49.92, 50.68, 71.14)\,{\rm meV}\,,\cr
(m_1,m_2,m_3)_{\rm IH}~&=(49.33, 50.10,~\,3.50)\,{\rm meV}.
}
In particular, these examples agree with the values of Ref.\,\cite{valle.12} within
1-$\sigma$.
The large departure from the minimal texture for the charged lepton mass matrix can
be seen from the fact that these points correspond $\beta_l\approx 0.928$ and
$\beta_l\approx 0.982$ for the QD and IH cases, respectively. Therefore, $|B_l|\gg
|B_l'|$ in \eqref{Ml}.
In this case, we lose some naturality in view of the structure \eqref{yukawa:l}.
For the case of IH, although $\beta_{\nu}$ is small, it cannot be much larger
[e.g. $\mathcal{O}(1)$], otherwise we lose the appropriate root of \eqref{eq:dc} or
\eqref{eq:c}.
It should be emphasized that a large set of solutions with parameters
$\{\beta_{0l},\beta_\nu,a,\bar{b},c\}$ similar to \eqref{qd.ih} can be easily found
and \eqref{qd.ih} is not special in this regard.
It is also evident that very different predictions arise from the extreme cases
\eqref{qd.ih}. For example, the effective mass that enters the neutrinoless double
beta decay experiments is quite large for these cases:
\eq{
\text{(QD) } m_{ee}\approx 50.1\,{\rm meV},\quad
\text{(IH) } m_{ee}\approx 48.1\,{\rm meV}.
}

As a remark, we should also note some similar analysis has been performed by
Ref.\,\cite{fritzsch:pert} on the nonhermitean departure from the Fritzsch
form, although it considers Dirac neutrinos.
Our modified form \eqref{M:mod} is a particular case of the NNI form considered
in Ref.\,\cite{fritzsch:pert} but the latter only analyzes the deviations
perturbatively.
The approximation \eqref{dc:approx} is better than the perturbative first-order
approximation.
Also, in our case, we have Majorana neutrinos.

%%%%%%%%%%%%%%%%%%%%%%%%%%%%%%%%%%%%%%%%%%%%%%%%%%%%%%%%%%%%%%%%
\section{Conclusions}
\label{sec:conclusions}

We have presented here a simple extension of SM based on the flavor group
$S_3$, which is capable of generating the minimal texture proposed in
Ref.\,\cite{yanagida} with three additional degrees of freedom:
(i) $M_R$ has two independent diagonal elements instead of one (we also have more
phases), (ii) the (23) and (32) elements of $M_l$ are independent
and (iii) the (23) and (32) elements of $M_D$ are independent as well.
If we equate the parameters in (i), (ii) and (iii) [and adjust the phases for (i)],
we reproduce exactly the ansatz of Ref.\,\cite{yanagida}.

The model possesses an additional flavor symmetry $G_F=S_3\otimes Z_5\otimes Z_3$
and five additional complex scalars $\zeta_{D}=(\zeta_1,\zeta_2),\eta,\chi,\chi'$.
The scalars $\chi,\chi'$ only couple to right-handed neutrinos through lepton
number violating interactions. In fact, $Z_3$ in $G_F$ can be extended to
$U(1)_{B-L}$ in the lepton sector if we assign $B-L=2$ to both $\chi,\chi'$.
This symmetry, however, is softly broken to $Z_3$ (factoring out the group
$\{\pm 1\}$ for leptons) by the term $\chi^3$ in the scalar
potential. Therefore, our flavor group could be a discrete remnant of a larger
symmetry containing $U(1)_{B-L}$ at higher energies.
We should emphasize that the symmetry $Z_5\otimes Z_3$ highly constrains the model,
naturally providing both a rationale for large tau mass and the necessary vacuum
alignment.

On the phenomenological side, the freedom (i) does not lead to large deviations
from the minimal texture.
The freedoms (ii) and (iii), on the other hand, can modify significantly the mixing
angles in the orthogonal matrices $O_l$ and $O_\nu$ that diagonalize the
respective squared mass matrices in the real form.
The freedoms (ii) and (iii) were parametrized by two parameters $\beta_{0\alpha}$
\eqref{def:beta}, $\alpha=l,\nu$, that may vary in the range
$|\beta_{0\alpha}|\lesssim 1$.
The values $\beta_{0l}=\beta_{0\nu}=0$ correspond to the minimal texture.
Since the dependence of the PMNS matrix on $\beta_{0l},\beta_{0\nu}$ is
smooth, the solutions for $\beta_{0l}=\beta_{0\nu}=0$\,\cite{yanagida.12}
are not disrupted as we relax $\beta_{0l}\approx\beta_{0\nu}\approx 0$.
However, as we deviate from the minimal texture, very different solutions are
possible. This was shown through various examples.
In the first example, we have shown in Fig\,\ref{fig2} that the minimal case can be
distinguished from the case $(\beta_{0l},\beta_{\nu})=(1,-0.418)$ with similar
values for the rest of the parameters.
For even larger deviations from the minimal texture, we have shown that both
quasidegenerate spectrum and inverted hierarchy are possible.
This contrasts sharply with the minimal texture where neither of them is
possible\,\cite{yanagida.12}.

In summary, the modified texture \eqref{M:mod} arising from our model for the Dirac
mass matrices of charged leptons and neutrinos easily accommodates the present
data on the neutrinos sector but still allows a wide range of possibilities if we
permit large deviations from the minimal texture.

\acknowledgments

The work of A.G.D and C.C.N is partially supported by the Brazilian FAPESP and
CNPq. The work of A.C.B.M. is supported by CAPES.

%%%%%%%%%%%%%%%%%%%%%%%%%%%%%%%%%%%%%%%%%%%%%%%%%%%%%%%%%%%%%%%%
\appendix
%%%%%%%%%%%%%%%%%%%%%%%%%%%%%%%%%%%%%%%%%%%%%%%%%%%%%%%%%%%%%%%%
\section{Alignment of $\aver{\zeta_D}$}
\label{ap:pot}

We justify here the alignment $\aver{\zeta_D}=(0,u_2)^\tp$ assumed in
\eqref{vevs}\,\cite{endnote2}.
For this alignment, the existence of the scalar $\eta$ transforming as $\bs{1}'$ of
$S_3$ and its coupling to $\zeta_D$ are essential. It is exactly
$\aver{\eta}\neq0$ that breaks $S_3$ to $Z_3$, whereas $\aver{\zeta_D}$
breaks this remaining symmetry.
To be explicit, we can write the generators $a,b$ of $S_3$ acting on $\zeta_D$
as\,\cite{ishimori}
\eq{
D(a)=\diag(1,-1)\,,\quad
D(b)=\begin{pmatrix}
  -\frac{1}{2} & \frac{\sqrt{3}}{2} \cr
  -\frac{\sqrt{3}}{2} & -\frac{1}{2}
\end{pmatrix}\,.
}
Both are contained in $O(2)$: $D(a)$ is a reflection in the $\zeta_{2}$
direction and $D(b)$ is a $2\pi/3$ rotation in the plane.
The vev $\aver{\eta}\neq0$ breaks the $Z_2$ generated by $D(a)$, and $S_3$ is broken
to the $Z_3$ generated by $D(b)$.

To analyze the alignment of $\zeta_D$, we consider only the relevant terms that
depend on $\zeta_D$, which are
\eq{
\label{V}
V_{\zeta_D} = \mu^2_d\zeta_D^\dag\zeta_D +\lambda_1(\zeta_D^\dag\zeta_D)^2
 +\lambda_2 \big([\zeta_D^*\zeta_D]_{1'}\big)^2
 +\ \big\{
  \lambda_3[\zeta_D\zeta_D\zeta_D]_{1^\prime}\eta^* +
  \lambda_3'[\zeta_D\zeta_D]_1\chi\chi^{\prime*}
  + h.c.\big\}
\,,
}
where we assume for our purposes that $\lambda_3,\lambda_3'$ are real; the
coefficient $\mu_d^2$
effectively includes terms depending on other $G_F$ invariant combination of
fields such as $\phi^\dag\phi$ or $\chi^*\chi$.
The term $([\zeta_D^*\zeta_D]_2)^2$ does not introduce new terms. We note that the first two terms  of the potential \eqref{V} are $U(2)$ invariant, and we
obviously need $\mu_d^2<0$ to obtain nontrivial $\aver{\zeta_D}$.
The third, fourth and fifth terms break the $U(2)$ invariance because
\alig{
\label{DD:1'}
([\zeta_D^*\zeta_D]_{1'})^2&=-4\im^2(\zeta_{1}^*\zeta_{2})\,,\\
\label{DDD:1'}
[\zeta_D\zeta_D\zeta_D]_{1'}&=
\zeta_{2}(\zeta_{2}^2-3\zeta_{1}^2)\,,\\
\label{DD:1}
[\zeta_D\zeta_D]_{1}&=
\zeta_{1}^2+\zeta_{2}^2\,.
}
The term \eqref{DD:1'} is still invariant by $O(2)$ and global rephasing.
The $O(2)$ symmetry is reduced to $S_3$ by the term proportional to \eqref{DDD:1'},
which is further reduced to $Z_3$ when $\eta$ acquires a nonzero vev.

To ensure the term proportional to \eqref{DD:1'} is positive semidefinite,
we choose $\lambda_2<0$. This choice tends to align the phases of $\aver{\zeta_{1}}$
and $\aver{\zeta_{2}}$, or otherwise make either $\aver{\zeta_{1}}=0$ or
$\aver{\zeta_{2}}=0$.
The term \eqref{DDD:1'} tends to make $\aver{\zeta_{1}}=0$.
One can then check that the direction $\aver{\zeta_D}=(0,u_2)^\tp$ is a
local minimum in the shell (orbit) where $\zeta_D^\dag\zeta_D$ is constant, provided
that $\aver{\eta}\neq 0$.
This check is most easily performed for the case where all the coefficients of
the scalar potential are real and the vevs of $\eta,\chi,\chi'$ are real as well.
Then we can choose the sign of the coefficients of \eqref{V} such that
\eq{
\lambda_3\aver{\eta}<0,\quad
\lambda_3'\aver{\chi{\chi'}^*}<0\,.
}
In this case, the $u_2$ corresponding to the deepest minimum in this direction will
be real and positive.

We should remark that due to the discrete $G_F$ symmetry, the potential \eqref{V}
possesses multiple degenerate minima.
After $\eta,\chi,\chi'$ have acquired nonzero vevs, the $Z_3\subset S_3$
symmetry guarantee that if $\aver{\zeta_D}=(0,u_2)^\tp$ is a minimum, then
\eq{
D(b)\aver{\zeta_D},\quad
D^2(b)\aver{\zeta_D},
}
are also degenerate minima. These minima are not aligned as $(0,u_2)$ but they
induce equivalent mass matrices since the form \eqref{Ml} is recovered after we
apply $D^{-1}(b)$ or $D^{-2}(b)$ transformations on the fermions transforming as
$\bs{2}$ of $S_3$.

Besides the last two terms in \eqref{V} there is still one last  nonhermitian term
invariant under $G_F$, which is contained  in the  potential involving only $\chi$,
\eq{
\label{Vchi}
V_{\chi} = \mu^2_\chi\chi^\dag\chi +\lambda_\chi(\chi^\dag\chi)^2
 +\ \big\{
  m\,\chi^3
  + h.c.\big\}
\,.
}
The remaining renormalizable terms in the total potential are all hermitian. We can
see there are parameter ranges of the potential that guarantee nonzero and arbitrary
vevs; we keep $\aver{\phi}$ in the electroweak value and the rest of the vevs may
be pushed to higher energies.

It must be noted that we have a rephasing transformation
$\zeta_D\rightarrow e^{i\alpha}\zeta_D,\,\,\eta \rightarrow
e^{i3\alpha}\eta,\,\chi'\rightarrow e^{i2\alpha}\chi'$, with $\phi$ and $\chi$
transforming trivially, resulting in a $U_\alpha(1)$ continuous accidental
symmetry, which contains $Z_5$ of $G_F$.
This generates a Goldstone boson after scalar fields acquire vevs. The
Goldstone boson is given by the following combination
\eq{
\label{Gol}
G = \frac{u_2\im \zeta_2 + 3u_\eta\im
\eta+2u^{\prime}_s\im \chi'}{\sqrt{u_2^2+9u_\eta^2+4u^{\prime 2}_s}}
\,.
}
$G$ would couple with the SM fields through nonrenormalizable interactions  in
\eqref{yukawa:l}, and they are suppressed by the scale $\Lambda$. Also, the coupling
could be even suppressed if $u_2,\,u_\eta\ll u^{\prime}_s$.  The main interaction
is with $N_D$ in Eq.\,\eqref{L:MR} and so $G$ is harmless.

%%%%%%%%%%%%%%%%%%%%%%%%%%%%%%%%%%%%%%%%%%%%%%%%%%%%%%%%%%%%%%%%

%%%%%%%%%%%%%%%%%%%%%%%%%%%%%%%%

\begin{thebibliography}{99}
\bibitem{endnote0}
If allow order $1/\Lambda^2$ terms in \eqref{yukawa:l}, we have do add the
terms $\bar{L}_DE_D\phi\zeta_D\zeta_D,\bar{L}_DE_D\phi\chi\chi^{\prime*}$,
which induce nonzero terms in the (11), (22), (12) and (21) entries of $M_l$.
These contributions are negligible if
$|\aver{\zeta_D}|/\Lambda,|\aver{\chi}|/\Lambda,|\aver{\chi'}|/\Lambda\ll 1$.
\bibitem{endnote1}
Strictly speaking, we need to ensure that $|\Delta b^2|$ is not so large as to
make either $b$ or $b'$ go to zero.
\end{thebibliography}
\end{document}